# Improved Watermarking Scheme Using Decimal Sequences

Ashfaq N Shaik

**Abstract:** This paper presents watermarking algorithms using d-sequences so that the peak signal to noise ratio (PSNR) is maximized and the distortion introduced in the image due to the embedding is minimized. By exploiting the cross correlation property of decimal sequences, the concept of embedding more than one watermark in the same cover image is investigated.

**Introduction**

This work improves upon the work reported by Mandhani and Kak [6] on watermarking using decimal sequences [1-3,8] by using ideas of Malvar and Florencio [5]. Adaptive gain correction is performed in the embedding procedure so that the peak signal to noise ratio (PSNR) can be maximized. Gain analysis was performed so as to estimate the optimum gain necessary for robust watermarking.

**Basics of Spread Spectrum Watermarking**

A data signal to be watermarked may be modeled as a random vector. Let the elements of the data signal be denoted by $x \in R^N$, which will be assumed to be independent identically distributed (i.i.d) Gaussian random variables having a standard deviation of $\sigma_x$ and mean zero. A watermark is defined as a direct spread spectrum sequence $w$, which is a random vector pseudo-randomly generated in $w \in \{\pm 1\}^N$. The watermarked signal is $y = x + \delta w$, where $\delta$ is the watermark gain. A basic model of this scheme is given in Figure 1.

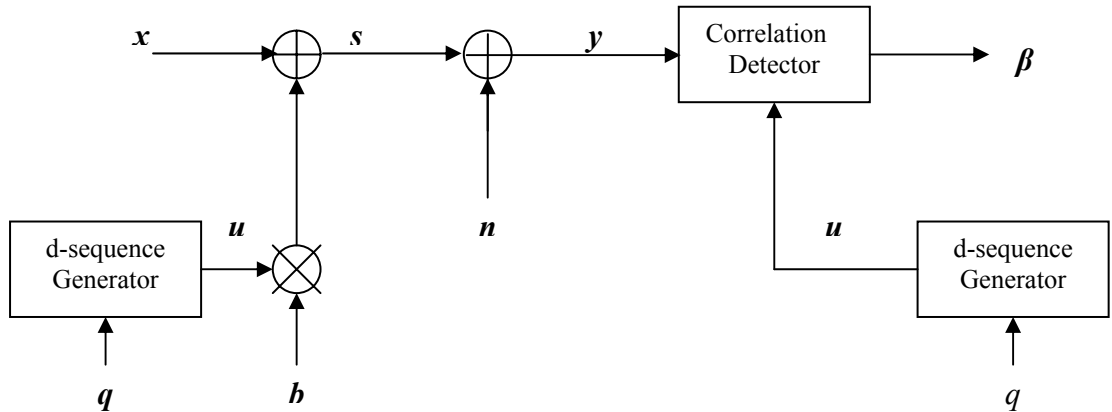

Figure 1 Realization of Decimal Sequence Spread Spectrum Watermarking.

The watermark bit $b \in \{-1, 1\}$ is multiplied by the pseudorandom sequence, and added to the cover object $x$. Watermark extraction is achieved by cross correlating with the pseudorandom sequence. The detection criteria is established using correlation analysis, that is the watermark $w$ is detected by correlating the given signal $y$ with $w$.

$$C(y, w) = y \cdot w = E[y \cdot w] + N\left(0, \frac{\sigma_x}{\sqrt{N}}\right)$$

If the signal $y$ has been marked and no malicious attacks or other signal modifications are performed, then E [$y \cdot w$] = δ, else E [$y \cdot w$] = 0. The threshold for detection is τ. The watermark is present if C ($y, w$) > τ. Under the conditions that $x$ and $w$ are i.i.d. signals, such a detection scheme is optimal. It is believed that spread spectrum (SS) method of watermarking can incorporate a high degree of robustness The major drawback of the SS watermarking scheme is that it requires a high gain value δ. To overcome this problem, Malvar, *et.al* [5] proposed the improved spread spectrum (ISS) technique. Here we use the ISS technique for decimal sequences to enhance the performance of the embedding procedure and improve the overall performance of the watermarking scheme.

**Decimal Sequence Image Watermarking**

The decimal sequence spread-spectrum watermarking scheme is shown in figure 1. The prime $q$ drives the decimal sequence (d-sequence) generator, produces the chip sequence $u$, which has zero mean and whose elements are equal of $-\sigma_u$ or $+\sigma_u$. The chip sequence $u$ is either added or subtracted from the signal $x$ depending on the value of the watermark bit $b$, which takes values $\{+1, -1\}$. The signal $s$ is the watermarked signal and $n$ is the noise introduced into the system.

A traditional analysis [5] of the watermarking scheme leads to the simple detection statistic. The inner product of two vectors $x, u$ is defined as follows:

$$\langle x, u \rangle = \frac{1}{N} \sum_{i=0}^{N-1} x_i u_i \quad and \quad \|x\| = \langle x, x \rangle \tag{1}$$

where N is the length of the vectors $x, s, u, n$ and $y$. The embedding is performed according to the following equation,

$$s = x + bu \tag{2}$$

Distortion $D$ associated with the embedded signal is defined as $\|s - x\|$.

$$D = \|s - x\| = \|bu\| = \|u\| = \sigma_u^2 \tag{3}$$

We assume that the channel is being modeled as an additive white Gaussian noise (AWGN). Thus,

$$y = s + n \tag{4}$$

Detection is performed based on the detection statistic $r$

$$r = \frac{\langle y, u \rangle}{\langle u, u \rangle} = \frac{\langle bu + x + n, u \rangle}{\sigma_u^2} = b + \bar{x} + \bar{n} \tag{5}$$

and the estimated bit

$$\beta = sign(r) \tag{6}$$

where

$$\bar{x} = \frac{\langle x, u \rangle}{\|u\|} \quad \& \quad \bar{n} = \frac{\langle n, u \rangle}{\|u\|} \tag{7}$$

We assume the simple statistical model for the signal $x$ and noise $n$. We assume these two to be Gaussian random processes. Therefore,

$$x_i \approx N(0, \sigma_x^2), \quad n_i \approx N(0, \sigma_x^2)$$

Thus the detection statistic $r$ is also Gaussian, i.e.

$$r \approx N(m_r, \sigma_r^2), \quad m_r = E[r] = b\sigma_r^2 = \frac{\sigma_x^2 + \sigma_n^2}{N\sigma_u^2} \tag{8}$$

**Improvement to Traditional Watermarking Scheme**

In this technique of watermarking we utilize the knowledge of the signal $x$ to modulate the energy of the inserted watermark so that the interference due to $x$ can be compensated for [5]. We define a function $g(\bar{x}, b)$ which helps in varying the amplitude of the chip sequence $u$. Thus we have

$$s = x + g(\bar{x}, b) \cdot u \tag{10}$$

For simplicity we assume that $g(\bar{x}, b)$ is a linear function. We can thus write

$$g(\bar{x}, b) = \mu b - \lambda \bar{x} \tag{11}$$

Thus,

$$s = x + \left(\mu b - \lambda \bar{x}\right) \cdot u \tag{12}$$

and the detection statistic would be

$$r = \frac{\langle s, u \rangle}{\|u\|} = \mu b + (1 - \lambda)\bar{x} + n \tag{13}$$

The closer we make λ to 1, the greater the influence of *x* is removed from *r*. The detector still remains the same, i.e., the detected bit is *sign(r)*. The expected distortion *D* is

$$E[D] = E[\|s - x\|]$$
$$= E[|\mu b - \lambda x|^2 \sigma_u^2] = \left(\mu^2 + \frac{\lambda^2 \sigma_x^2}{N\sigma_u^2}\right)\sigma_u^2 \tag{14}$$

In order to have the same distortion as the traditional distortion we force D to be equal to that of the traditional distortion (equation 3). We have

$$\left(\mu^2 + \frac{\lambda^2 \sigma_x^2}{N\sigma_u^2}\right)\sigma_u^2 = \sigma_u^2 \tag{15}$$

Or

$$\mu = \sqrt{\frac{N\sigma_u^2 - \lambda^2 \sigma_x^2}{N\sigma_u^2}} \tag{16}$$

We can choose a particular distortion value by equating Equation 15 to be equal to a fraction of the traditional distortion. We compute the mean and variance of the sufficient statistic. They are

$$m_r = \mu b, \quad \text{and} \tag{17}$$

$$\sigma_r^2 = \frac{\sigma_n^2 + (1-\lambda)^2 \sigma_x^2}{N\sigma_u^2} \tag{18}$$

The error probability can be calculated as follows:

$$p = \frac{1}{2} erfc\left(\frac{m_r}{\sigma_r \sqrt{2}}\right) \tag{19}$$

$$= \frac{1}{2} erfc\left(\frac{1}{\sqrt{2}} \sqrt{\frac{\frac{N\sigma_u^2}{\sigma_x^2} - \lambda^2}{\frac{\sigma_n^2}{\sigma_x^2} + (1-\lambda)^2}}\right) \tag{20}$$

The maximum value of error probability *p* can be found by equating $\frac{\partial p}{\partial \lambda} = 0$, we get the optimum value of λ.

$$\lambda_{opt} = \frac{1}{2}\left\{\left(1 + \frac{\sigma_n^2}{\sigma_x^2} + \frac{N\sigma_u^2}{\sigma_x^2}\right) - \sqrt{\left(1 + \frac{\sigma_n^2}{\sigma_x^2} + \frac{N\sigma_u^2}{\sigma_x^2}\right)^2 - 4\frac{N\sigma_u^2}{\sigma_x^2}}\right\} \tag{21}$$

According to equation 14, we can reduce the distortion in the improved scheme by selectively choosing the values of λ and μ. Let us assume that we want a distortion level 10 times lesser than that of the traditional level. Thus using equation 15 and equating it to one tenth of the distortion of the traditional scheme we get

$$\left(\mu^2 + \frac{\lambda^2 \sigma_x^2}{N\sigma_u^2}\right)\sigma_u^2 = \frac{\sigma_u^2}{10} \tag{22}$$

or

$$\mu = \sqrt{\frac{N\sigma_u^2 - 10\lambda^2\sigma_x^2}{10N\sigma_u^2}} \tag{23}$$

For a simplistic model let us initially consider that there is no channel noise associated with the scheme, i.e. $\sigma_n^2 = 0$, thus we can assume that

$$\lambda_{opt} = 1$$

Substituting this value of $\lambda_{opt}$ and taking $\sigma_u^2 = 1$ in equation 23 we get

$$\mu = \sqrt{\frac{N - 10\sigma_x^2}{10N}} \tag{24}$$

We utilize the above equations to watermark grayscale images. The cover image is a grayscale image of Lena, the value of $\sigma_x^2$ is nearly equal to 2045 with a standard deviation of 45.22, also N=512*512 (size of image) thus substituting these values in equation 24 we get

$$\mu \approx \sqrt{\frac{1}{10}} = 0.316$$

We choose this value of μ and λ for our embedding algorithm. From equation 7 we have $\bar{x}$ =0.0299 for the decimal sequence generated by the prime $q$=8069. Thus the equation 12 becomes

$$s = x + (0.316*b - 0.0299) \cdot u \tag{25}$$

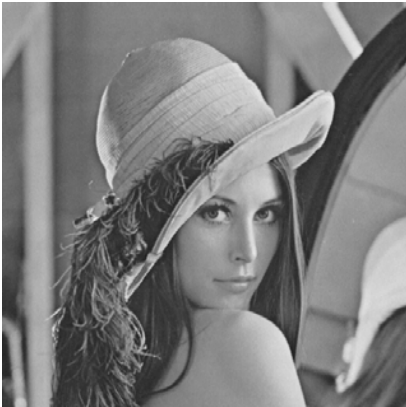 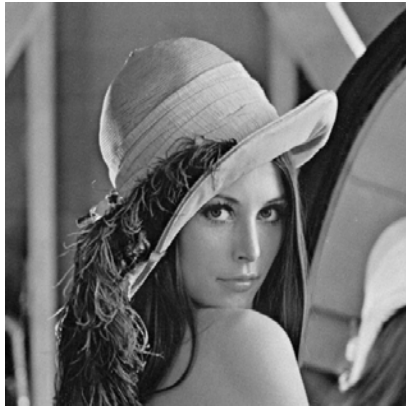

Figure 2 Watermarked Image, PSNR=36.05 dB.          Figure 3 Watermarked Image, PSNR =39.22 dB.

Now we chose another prime $q$=2467 to compare the results. From equation 7 we have $\bar{x}$ =0.0367.
Thus equation 25 changes to

$$s = x + (0.316*b - 0.0367) \cdot u \tag{26}$$

We observe from equations 25 and 26, that the variations in the second term are very small, which means that the effect of the cover image in the embedding procedure is almost negligible. The Peak Signal to Noise Ratio (PSNR) for this prime was found to be 39.22 dB.

Figure 4 Embedded Watermark.          Figure 5 Recovered Watermark.

The performance with other images was equally good.

**Gain Analysis of Decimal Sequence Watermarking Scheme**

Simulations were performed to analyze the effects of varying the parameter $\mu$ in the embedding algorithm. We refer this to be the gain. By optimizing the value of $\mu$ we can find the most effective scheme for embedding the watermark in the cover object. We performed simulations by varying $\mu$ from 0.1 to 1 in steps of 0.05. Primes were chosen at random for this simulation and the results are shown in figure 6. We observe that for most primes the optimal gain value lies close to 0.3, which indicates that with our proposed scheme using decimal sequences as the random sequences we can reduce the distortion *D* almost 10 fold when compared to the traditional watermarking scheme using pseudorandom sequences as the spreading sequences. Also for certain primes perfect detection is possible for gain values as low as 0.175. These primes can be used to further reduce the distortion. Certain primes have low gains associated with them. Primes, whose decimal sequence has period *p* equal to *(q-1)/n,* where *n* is even, exhibit good statistical properties. The autocorrelation of such sequences is lower when compared to that of primes whose periods are *(q-1)/n,* where *n* is odd. Whenever *n* is odd, the first *(p-1)/2* digits of the decimal sequence will be exactly the complement of the subsequent digits. The first *(p-1)/2* digits are sufficient to generate the last *(p-1)/2* digits. This is the reason why such primes have higher gain values. For any prime the optimal value of the gain should be greater than 0.32. Those primes which have gains lesser than 0.32 may be used to embed more information in the cover object because the distortion associated with these primes is lesser than that of primes whose gains are larger than 0.32.

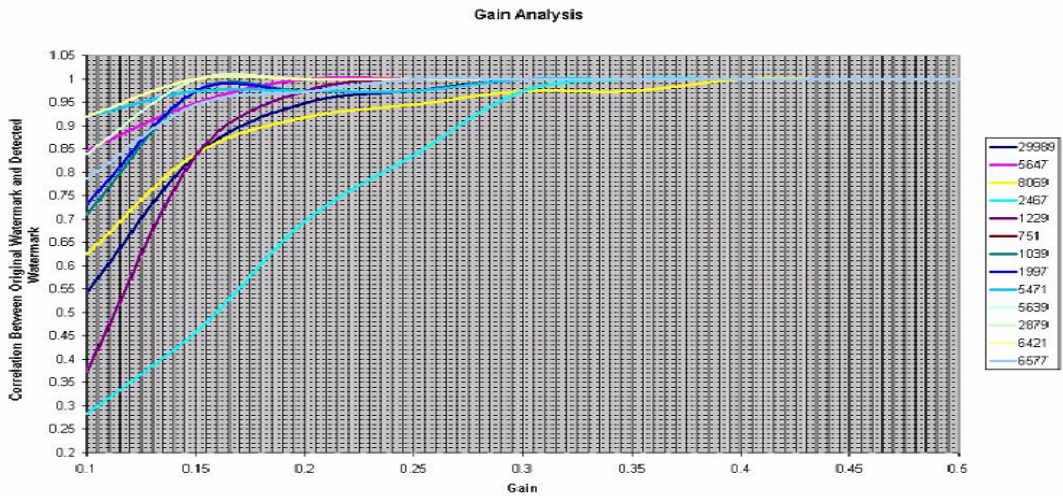

Figure 6 Gain Analysis Graph

Thus more information can be hidden using primes whose gains are less than 0.3 for the same distortion value *D*. For example for the prime *q=5647,* the period of the decimal sequence generated is 2823 and the optimal gain associated with this is about 0.2. Thus we get

$$\mu \approx 0.2 = \sqrt{\frac{1}{x}}$$

Solving the above equation for *x*, we get *x=25*. The distortion can be reduced by a factor of 25, when compared to the traditional method.

**Multiple Watermark Embedding Using Decimal Sequences**

As discussed in previous sections decimal sequences can be used with very low gain values, this property hints that we can perhaps embed multiple watermarks in the same cover object without causing considerable distortion in the cover image. The cross correlation properties of decimal sequences [1,2] are good, since the cross correlation function of two maximum length decimal sequences in the symmetric form is identically equal to zero if the ratio $k_1/k_2$ of their periods reduces to a irreducible fraction $n_1/n_2$, where either $n_1$ or $n_2$ is an even number. By choosing the prime $q_1$ and $q_2$ we can generate decimal sequences whose cross-correlation is equal to zero. We used the primes 2467 and 8069 with periods 2466 and 4019 respectively. $\bar{x}$ for $q=2467$ was found to be 0.0367 and that for $q=8069$ was –0.0299. First we embedded the watermark shown in figure 7, the PSNR was found to be 37.78 dB. Next we embedded the second watermark as shown in figure 8, the PSNR reduced to 34.98.

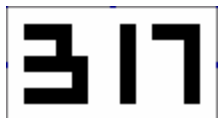
Figure 7 Watermark 1.

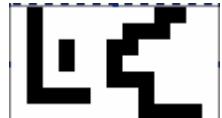
Figure 8 Watermark 2.

The detection for both the primes 2467 and 8069 was perfect. This indicates to us that we can embed more than one watermark in the same cover object with some deterioration in the PSNR. The detection statistics are shown in figures 9 and 10 for the primes 2467 and 8069 respectively.

We experimented with various combinations of primes and found that in all combinations the detection of the watermarks was perfect. Thus we have not only improved the previous decimal sequence watermarking scheme but have also provided a way of embedding more than one watermark in the same cover object.

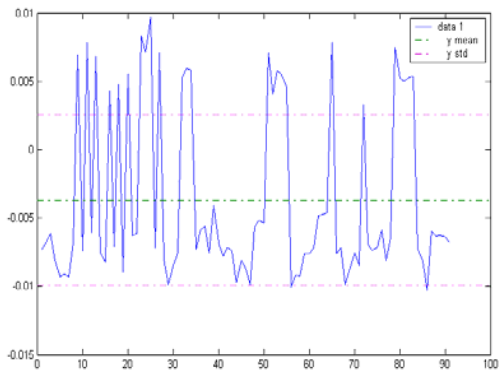
Figure 9 Detection Statistic for Prime 2467.

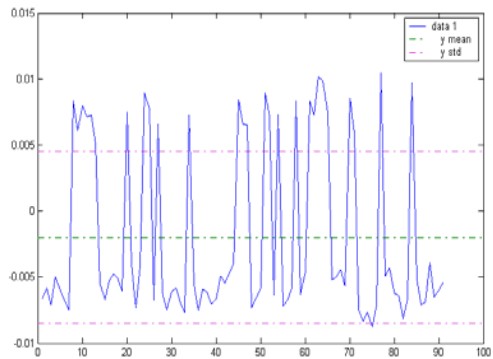
Figure 10 Detection Statistic for Prime 8069.

**Analysis of Results**

We preformed simulations for various images and various primes and chose the values of $\mu = 0.316$ and $\lambda = 1$ for our simulations. we were able to recover the watermarks perfectly for different cover images and also for different watermarks. As shown in Figure 6 the most optimal value of gain can be as low as 0.175 and for all values of gain over 0.4 the recovery of the watermark was always perfect. This fall in gain was possible because of the watermarking algorithm. Our algorithm provides the flexibility of considering the effects of noise also. Also the distortion introduced in the cover image can be adjusted appropriately using these values of $\mu$ and $\lambda$. The values of $\mu$ and $\lambda$ can be appropriately chosen so that the effect of the cover image can be reduced in the embedding algorithm.

**Conclusions**

By appropriately selecting the primes and the gain associated with the embedding procedure we can reduce the distortion introduced in the cover image. Our proposed scheme has the following advantages over the earlier watermarking scheme:
- The noise immunity of the watermarked image is increased by appropriately choosing the prime and the values of $\mu$ and $\lambda$.
- We have reduced the distortion introduced in the image. We have been able to use gains as low as 0.175.
- Multiple watermarks may be embedded using different decimal sequences by exploiting the cross correlation property.

- Using decimal sequences as compared to pseudorandom sequences we were able to embed almost twice the data in a cover image for a fixed distortion value D. This is due to the fact that in pseudorandom sequence based systems, complete correlation is not possible because of the large periods of the pseudorandom sequences. Hence to compensate for this, the gain has to be increased considerably.

This paper is limited to watermarking of grayscale images in the spatial domain. The watermark is a black and white image. Further research can be carried out for color image watermarking, video watermarking and also audio watermarking. If one were to embed a watermark with more than two symbols, advanced algorithms for embedding and detection can be devised. We may use different decimal sequences to embed different symbols of a watermark. Further research should include the use of other d-sequence related random sequences [4,7].

**References**

1. Kak, S., "Encryption and error-correction coding using d-sequences," IEEE Trans. on Computers, vol. C-34, pp. 803-809, 1985.
2. Kak, S., and Chatterjee, A., "On decimal sequences," IEEE Trans. on Information Theory, vol. IT-27, pp. 647-652, 1981.
3. Kak, S., "New results on d-sequences," Electronics Letters, vol. 23, pp. 617, 1987.
4. Kak, S., "The cubic public-key transformation," 2006. arXiv: cs.CR/0602097
5. Malvar, H.S., and Florêncio A.F.D, "Improved spread spectrum: A new modulation technique for robust watermarking," IEEE Trans. on Signal Processing, vol. 51, Apr 2003.
6. Mandhani, N., and Kak, S., "Watermarking using decimal sequences", Cryptologia, vol 29, pp. 50-58, Jan 2005; arXiv: cs.CR/0602003
7. Parakh, A., "A d-Sequence based recursive random number generator," 2006. arXiv: cs.CR/0603029
8. Vaddiraja, R., "Generalized d-sequences and their applications in CDMA Systems," MS Thesis, Louisiana State University, 2003.